\documentclass[journal]{IEEEtran}
\usepackage{enumitem}
\usepackage{xspace}
\usepackage{amsmath,amsfonts}
\usepackage{algorithmic}
\usepackage{algorithm}
\usepackage{array}
\usepackage[caption=false,font=normalsize,labelfont=sf,textfont=sf]{subfig}
\usepackage{textcomp}
\usepackage{stfloats}
\usepackage{url}
\usepackage{verbatim}
\usepackage{graphicx}
\usepackage{cite}
\usepackage{xcolor}
\usepackage[colorlinks=true, linkcolor=blue, citecolor=blue, urlcolor=black]{hyperref}
\usepackage{orcidlink}
\usepackage{booktabs}
\usepackage{multirow}
\usepackage{makecell}
\usepackage{cleveref}
\crefname{figure}{Fig.}{Figs.}
\usepackage{listings}

\def\BibTeX{{\rm B\kern-.05em{\sc i\kern-.025em b}\kern-.08em
    T\kern-.1667em\lower.7ex\hbox{E}\kern-.125emX}}
\usepackage{balance}

\newcommand{\TITLE}{\textsc{Ancora}\xspace}

\begin{document}

\title{\TITLE: Accurate Intrusion Recovery for Web Applications}

\author{Yihao~Peng$^{\orcidlink{0000-0002-9190-531X}}$\textsuperscript{\hyperlink{equal}{\textdagger}},~\IEEEmembership{Graduate~Student~Member,~IEEE},~Biao~Ma$^{\orcidlink{0009-0001-9372-1020}}$\textsuperscript{\hyperlink{equal}{\textdagger}},\\~Hai~Wan$^{\orcidlink{0000-0002-9608-5808}}$,~Xibin~Zhao$^{\orcidlink{0000-0002-6168-7016}}$,~\IEEEmembership{Senior~Member,~IEEE}

\thanks{Yihao Peng, Biao Ma, Hai Wan, and Xibin Zhao are with the Beijing National Research Center for Information Science and Technology (BNRist), Key Laboratory for Information System Security, Ministry of Education (KLISS), School of Software, Tsinghua University, Beijing 100084, China (e-mail: \url{xy20130630@gmail.com}; \url{mb_thu@163.com}; \url{wanhai@tsinghua.edu.cn}; \url{zxb@tsinghua.edu.cn}).}

\thanks{\protect\hypertarget{equal}{\textsuperscript{\textdagger}These authors contributed equally to this work.}}}

\markboth{IEEE TRANSACTIONS ON DEPENDABLE AND SECURE COMPUTING,~Vol.~, 2025}{Peng and Ma \MakeLowercase{\textit{et al.}}: \TITLE: Accurate Intrusion Recovery for Web Applications}


\maketitle

\begin{abstract}
Modern web application recovery presents a critical dilemma. Coarse-grained snapshot rollbacks cause unacceptable data loss for legitimate users. However, surgically removing an attack's impact is hindered by a fundamental challenge in high-concurrency environments: it is 
difficult to attribute the resulting file, database modifications, etc., to a specific attack related request. 
We present \TITLE, a system for precise intrusion recovery in web applications without invasive instrumentation. Our approach first isolates the full sequence of syscalls triggered by a single malicious request. Based on this sequence, \TITLE addresses file and database modifications separately. To trace file modifications, it builds a provenance graph that reveals all changes, even those made by exploit-spawned processes. To attribute database operations, a more difficult challenge due to connection pooling, \TITLE introduces a novel technique based on a spatiotemporal anchor. This anchor uses the request's network connection tuple and its active time window to pinpoint the exact database operations.
With all malicious file and database operations precisely identified, \TITLE performs a unified rewind and selective replay recovery. It first reverts the system to a clean snapshot taken before the attack. Then, it selectively re-applies only the legitimate operations to both the file system and the database. This process completely removes the attack's effects while preserving all concurrent legitimate data.
We evaluated \TITLE on 10 web applications and 20 CVE-based attack scenarios with concurrency up to 150 connections. Our experiments demonstrate that \TITLE achieves 99.9\% recovery accuracy with manageable overhead: up to 19.8\% response latency increase and 17.8\% QPS decrease in worst-case scenarios, and recovery throughput of 110.7 database operations per second and 27.2 affected files per second, effectively preserving legitimate data.

\end{abstract}


\section{Introduction}

\IEEEPARstart{W}{eb} applications have become primary targets for cyberattacks \cite{noauthor_owasp_nodate,shahid2022comparative}. While intrusion detection systems (IDS) offer a first line of defense \cite{cheng2024kairos,zengy2022shadewatcher,wang2022threatrace,wang2020youprovdetector,goyal2024rr-caid,han2020unicorn,xiong2020conan,han2021sigl}, they are not infallible. Consequently, the ability to precisely recover from successful intrusions to ensure data integrity and service continuity is paramount. Modern applications manage distributed and heterogeneous state, encompassing files scattered across disks, relational and non-relational databases, and external web services \cite{matos2021sanare,ammann2002recovery,akkucs2010data,chandra2011intrusion,chandra2013asynchronous,nascimento2015shuttle,matos2017rectify}. Handling any external request (including attack requests) may involve file operations, database operations, and interactions with external web services.

Industrial solutions typically rely on snapshot-based rollbacks, which indiscriminately discard all post-attack operations, including legitimate user transactions, leading to unacceptable data loss \cite{veeam,zerto,rubrik,cohesity}. 
A more effective paradigm is ``surgical'' recovery, which precisely analyzes and reverses only malicious changes. \textit{The central task of such recovery is to identify all modifications—spanning files, databases, and external web services—caused by a single malicious request.} However, achieving this in high-concurrency environments is exceedingly challenging. Interleaved streams of low-level syscalls and database logs from numerous concurrent users create an ambiguous many-to-many mapping, severely complicating the attribution of specific database operations and file modifications to a single originating http request.

\smallskip
\noindent\textbf{Previous Solutions and Limitations.} Prior research has explored two main avenues to address this challenge.

\begin{itemize}[leftmargin=*]
    \item \textbf{Taint analysis based on metadata propagation} instruments an application's runtime, such as the language interpreter or middleware, to propagate a unique identifier from each http request to its subsequent operations  \cite{ammann2002recovery,akkucs2010data,chandra2011intrusion,chandra2013asynchronous,nascimento2015shuttle}. This allows for direct correlation. However, this approach is fundamentally limited by its high cost and invasiveness, requiring deep, complex modifications to application frameworks that are impractical in production environments. Furthermore, it is inherently blind to unexpected execution paths created by exploits, such as remote command execution, because these actions bypass the instrumented logic and thus remain untainted and untraceable.

    \item \textbf{Black-box behavior pattern matching} avoids instrumentation by using external monitoring and machine learning to infer correlations between high-level http requests and low-level system events  \cite{matos2017rectify,matos2021sanare}. The main limitation of this approach is its diminished accuracy in high-concurrency settings. The dense, interleaved event streams make precise attribution challenging for learning-based models, often resulting in misclassifications. These systems are also typically restricted to monitoring predefined directories for performance reasons, leaving them unable to detect malicious file writes in arbitrary, unpredictable locations.
\end{itemize}

\noindent\textbf{Our Solution.}
This paper presents \TITLE, a system for precise intrusion recovery in web applications. \TITLE comprehensively traces and reverses all system state changes, across both files and databases, caused by a malicious request. It achieves this with only application-agnostic, framework-level instrumentation.

Inspired by attack investigation techniques from forensics analysis, which trace the causal chain of events forward from an initial point of compromise \cite{wang2023tesec,lee2013high,yu2021alchemist,ma2017mpi,alhanahnah2022autompi,hassan2020omegalog}, \TITLE begins by using forward analysis to isolate the complete syscall sequence corresponding to a given malicious http request. This sequence alone is sufficient to identify and report all interactions with external web services. However, recovering the internal state requires solving two further challenges at a higher semantic level:

\begin{enumerate}[leftmargin=*]
    \item Database Correlation Challenge: Accurately linking an http request to its specific database operations through complex layers of application connection pooling and asynchronous database processing.
    \item File Correlation Challenge: Tracing covert file writes that bypass normal application logic, often triggered by vulnerabilities like command injection.
\end{enumerate}

{To solve the database challenge}, we deconstruct the complex application-database interaction model by using network communication metadata as a \textit{spatiotemporal anchor}. From the request's syscall sequence, \TITLE extracts the unique network connection tuple (4-tuple) and its precise active time window. This anchor allows us to pinpoint the exact database server thread that handled the request and then filter its I/O on database log files, thereby isolating the exact database operations belonging to the malicious request.

{To solve the file challenge}, we employ \textit{causality tracking based on syscalls} \cite{li2021threat,hossain2018dependence}. Instead of passively monitoring a few directories, \TITLE actively analyzes the request's syscalls to construct a complete provenance graph. This graph reveals the entire causal chain of process execution triggered by the initial malicious act, making any covert file writes by descendant processes visible, regardless of their location on the file system.

With all malicious operations identified, \TITLE executes a unified \textit{Rewind \& Selective Replay} recovery. It first rolls the system back to a clean snapshot. Then, it performs a fine-grained replay: for the database, it replays only legitimate transactions from the database's own logs; for the file system, it uses a hybrid strategy, restoring critical system areas from baseline backups while incrementally replaying legitimate writes in user data directories. This process eradicates the attack's impact while preserving all concurrent legitimate user data.

\smallskip
\noindent\textbf{Experiments and Evaluations.} We evaluated \TITLE on 10 web applications spanning 4 programming languages. The experiments included 20 real CVE vulnerabilities and covered three common SQL and NoSQL databases: \texttt{MySQL}, \texttt{PostgreSQL}, and \texttt{MongoDB}. The results show that \TITLE achieves 99.9\% accuracy under concurrency up to 150 connections with manageable overhead: up to 19.8\% response latency increase and 17.8\% QPS decrease in worst-case scenarios, and recovery throughput of 110.7 database operations per second and 27.2 affected files per second.

\smallskip
\noindent\textbf{Our Contributions:}
\begin{itemize}[leftmargin=*]
    \item We design and implement \TITLE, a novel recovery system that uses causal tracing of low-level behaviors to precisely restore file and database state and identify external interactions, requiring only non-intrusive, framework-level instrumentation.
    \item We propose a database operation tracing method that deconstructs the asynchronous execution model using network metadata as an anchor, solving the correlation problem in the presence of connection pools and multi-threading.
    \item We design a file system repair mechanism that combines causality tracking with a hybrid rewind-and-replay strategy, enabling reliable recovery of file tampering at any path by discovering covert writes through a provenance graph.
    \item We demonstrate through extensive experiments on diverse applications and attack scenarios that \TITLE achieves 99.9\% accuracy and incurs manageable performance overhead with response latency increase up to 19.8\% and QPS decrease up to 17.8\% in worst-case scenarios under typical workloads.
\end{itemize}

In the spirit of open science, we make our tool available at \url{https://github.com/ma-b21/Ancora}.

\section{Background and Motivation}

\subsection{Provenance Graph}
A provenance graph is a directed acyclic graph that captures the causal and informational dependencies among system entities \cite{li2021threat}. Nodes in the graph represent subjects, such as processes and threads, and objects, such as files and network sockets. Directed edges represent the information flow and causal relationships between them. Every low-level system event, such as a process \textit{u} writing to a file \textit{v} at time \textit{t}, is abstracted as an edge in the graph \cite{xu2022depcomm}. When the object of one event becomes the subject of a subsequent event, a causal chain is formed, providing a theoretical foundation for tracing complex behavioral sequences. 
\textit{Forward analysis} on a provenance graph interprets these causal relationships to understand an attack's impact \cite{hossain2018dependence,zhang2016causality}. It begins from a known malicious source and traverses all subsequent events triggered by it, thereby assessing the full scope of the compromise.

\TITLE leverages this capability to address the challenge of covert file tampering caused by exploits. When a malicious http request is identified, \TITLE uses its corresponding initial process or thread as the starting point for analysis. By performing forward analysis, it constructs a complete causal execution chain. This chain exposes all file write operations performed by the malicious request and any of its descendant processes, regardless of their location. This provides the critical foundation for \TITLE's comprehensive attack footprint identification and precise file system recovery.

\subsection{Threat Model}
We consider an external attacker who sends one or more malicious http requests to exploit vulnerabilities in a web application, such as SQL injection, remote command execution, or arbitrary file uploads. The immediate goal of these attacks is to tamper with the system's internal state, specifically by making unauthorized modifications to the file system (e.g., writing a webshell, altering configuration files) or performing malicious database operations (e.g., stealing, modifying, or deleting data). The core objective of \TITLE is to precisely recover all file and database state changes caused by such malicious requests after they are detected.

\noindent\textbf{Assumptions.} Our threat model relies on the following assumptions:

\begin{itemize}[leftmargin=*]
\item The operating systems of the web and database servers provide reliable audit logging that records system-level events (i.e., syscalls), including process/thread identifiers, timestamps, and specific operations.
\item The audit logging mechanism and the \TITLE system are part of the Trusted Computing Base (TCB). We assume their integrity and availability are maintained during an attack and that they cannot be easily tampered with or disabled. This is a common and necessary assumption for many log-based security analysis systems. \cite{pohly2012hi,bates2015trustworthy,pasquier2017practical}.
\item The target database can be configured to enable comprehensive logging of modification statements, such as setting \texttt{log\_statement = 'mod'} in \texttt{PostgreSQL}.
This assumption, shared with other database recovery research \cite{matos2021sanare,nascimento2015shuttle,matos2017rectify}, ensures that the raw text of all database operations is captured, providing the necessary data for subsequent analysis and recovery.
\item The database environment possesses common and well-established mechanisms for creating snapshots and performing restorations (e.g., \texttt{mysqldump} for \texttt{MySQL}, \texttt{pg\_dump/pg\_restore} for \texttt{PostgreSQL} \cite{pgdump,pgrestore,mysqldump}), providing the essential basis for system state regression and the assurance of data consistency.
\end{itemize}

\noindent\textbf{Scope.} \TITLE focuses on attacks initiated by external requests that produce observable state changes at the system level. Our model does not directly address attacks that operate purely within the application's logic without triggering discernible differences in database or file I/O at the syscall level. Furthermore, the following attack vectors are considered out of scope: side-channel attacks, hardware-level exploits (e.g., Spectre), resource-exhaustion denial-of-service attacks (e.g., network floods), passive remote system fingerprinting, various network spoofing attacks, and any threats requiring prior physical access or an established foothold within the internal network. This threat model aligns with the consensus in existing web application recovery research \cite{matos2021sanare,matos2017rectify,nascimento2015shuttle}, allowing us to concentrate on the core challenge of reversing multi-faceted state corruption caused by external attacks.

\section{System Design}

\subsection{System Design Overview}

\begin{figure*}[!t]
    \centering
    \includegraphics[trim=0.5cm 0.2cm 0.5cm 0.2cm, clip,width=0.9\linewidth]{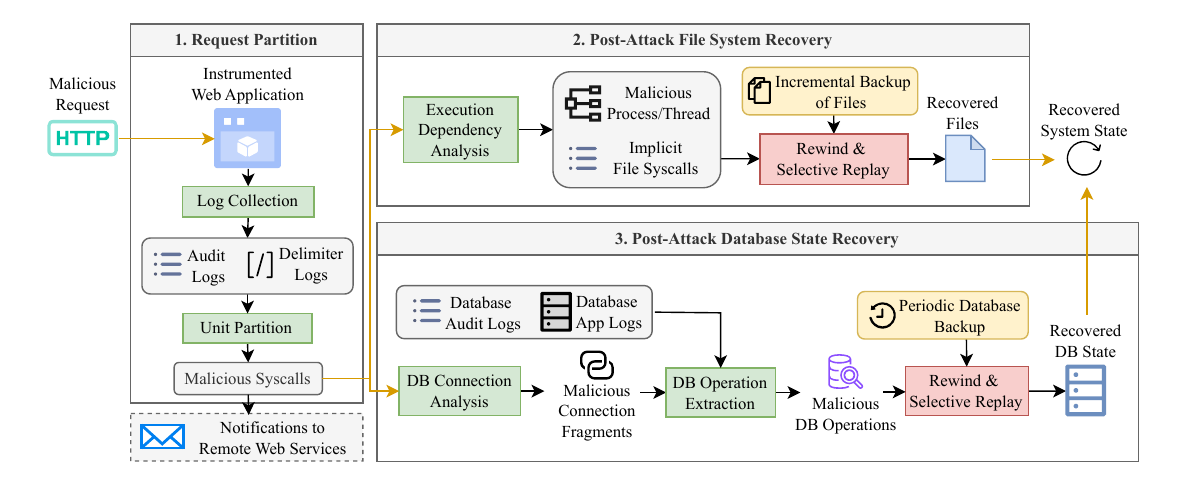}
    \caption{The workflow of \TITLE. Starting from a detected malicious http request, \TITLE first partitions syscall logs to isolate the request’s syscall sequence, then performs provenance-based file recovery and spatiotemporal-anchor-based database recovery, and finally conducts unified rewind and selective replay to remove all malicious effects while preserving concurrent legitimate data.}
    \label{fig:overview}
\end{figure*}

The core objective of \TITLE{} is to achieve comprehensive and accurate recovery by precisely identifying and reversing all state changes caused by a malicious request, across both files and databases, while preserving all concurrent legitimate operations. As illustrated in \cref{fig:overview}, the \TITLE{} workflow proceeds in three stages after a malicious http request is identified by an external IDS or a human analyst:

\begin{enumerate}[leftmargin=*]
    \item \textbf{Request Partition.} This foundational stage isolates the actions of a single malicious request. The \textit{Log Collection} module gathers syscall audit logs from the instrumented web application. We use lightweight, framework-level instrumentation to inject special \textit{Delimiter Logs}, each with a unique request ID, into the log stream. The \textit{Unit Partition} module then uses these delimiters to deterministically extract the exact sequence of \textit{Malicious Syscalls} corresponding to the malicious request from the interleaved system-wide log. This sequence serves as the input for the next stages and directly reveals any interactions with external services, enabling the generation of \textit{Notifications to Remote Web Services}. In parallel, \TITLE{} maintains data baselines through \textit{Incremental Backup of Files} and \textit{Periodic Database Backup}.

    \item \textbf{Post-Attack File System Recovery.} This stage restores the file system. It takes the \textit{Malicious Syscalls} as input. Instead of passively scanning directories, the \textit{Execution Dependency Analysis} module actively builds a provenance graph from the syscalls. This process traces the full causal chain, revealing all \textit{Implicit File Syscalls} triggered by the initial request or any of its descendant \textit{Malicious Processes/Threads}, thus exposing covert writes at any location. The \textit{Rewind \& Selective Replay} module then uses this precise footprint, along with the \textit{Incremental Backup of Files}, to perform a surgical recovery, yielding the \textit{Recovered Files}.

    \item \textbf{Post-Attack Database State Recovery.} This stage restores the database. Starting with the same \textit{Malicious Syscalls}, the \textit{DB Connection Analysis} module extracts the ``spatiotemporal anchor'' of the request's database interaction: its network connection tuple and precise active time window. This produces \textit{Malicious Connection Fragments}. The \textit{DB Operation Extraction} module then uses this anchor to pinpoint and extract the exact \textit{Malicious DB Operations} from the database's own logs (\textit{Database Audit Logs} and \textit{Database App Logs}). Finally, the \textit{Rewind \& Selective Replay} module uses these operations and a \textit{Periodic Database Backup} to roll back and selectively replay only legitimate transactions, producing the \textit{Recovered DB State}.
\end{enumerate}

Ultimately, this two-pronged recovery restores the system to a clean and consistent \textit{Recovered System State}, eradicating the attack's impact while maximizing the preservation of legitimate data. We detail the methods for each stage in Sections~\ref{sec:partition}, \ref{sec:file_recovery}, and \ref{sec:db_recovery}.

\subsection{Request Partition}
\label{sec:partition}

The prerequisite for precise recovery in \TITLE{} is the ability to isolate the complete, independent causal chain of system activities triggered by a single malicious http request from a noisy, interleaved stream of low-level system behaviors. In a production environment, syscall logs from the web server, database, and other system processes are voluminous and highly interleaved, which makes attributing low-level behaviors to a specific high-level request a significant challenge.

To address this, the \textit{Request Partition} module deterministically partitions the commingled stream of system-wide syscall audit logs into discrete behavioral units, each corresponding to a single http request. This module adopts a delimiter-logging approach, inspired by prior work on syscall partitioning \cite{wang2023tesec}, which uses lightweight, application-agnostic instrumentation at the web framework level. At critical points in a request's lifecycle, such as its initiation and context switches, our instrumentation injects special \textit{delimiter logs} containing a unique request ID into the syscall stream. These delimiters act as markers, delineating the boundaries of different requests' activities within the continuous log data.

\TITLE{} adapts its parsing strategy based on the web server's concurrency model:

\begin{itemize}[leftmargin=*]
    \item \textbf{Process/Thread-based Concurrency.} For servers that assign a dedicated process or thread to each request, such as certain modes of \texttt{Gunicorn}, the process ID (PID) and thread ID (TID) in syscall logs provide a natural grouping. Within a single PID/TID context, requests are handled sequentially. Therefore, our injected delimiters clearly mark the boundary points where the processing of one request ends and the next begins.

    \item \textbf{Asynchronous/Coroutine-based Concurrency.} For asynchronous models like those used by \texttt{FastAPI} and \texttt{Node.js}, multiple coroutines handling different requests execute concurrently within the same system thread. Their syscalls become deeply interleaved, making PID/TID insufficient for separation. In this scenario, our instrumentation is more fine-grained, injecting delimiters not only at the beginning of a request but also at every coroutine context switch, when one request's execution is suspended and another's resumes. These frequent, ID-tagged delimiters allow us to accurately unravel the intertwined syscalls from the single thread's log stream and re-attribute them to their respective source requests.
\end{itemize}

Using this approach, the \textit{Request Partition} module efficiently and accurately extracts the dedicated syscall sequence for every incoming malicious request. This sequence is the cornerstone of all subsequent analysis in \TITLE{}: it serves as the input for the file system and database recovery modules, and it also directly exposes the request's interactions with external web services (manifested as network-related syscalls like \texttt{sendto}), providing precise data for generating alerts and enabling cross-system remediation measures.

\subsection{Post-Attack File System Recovery}
\label{sec:file_recovery}

Once the malicious request's syscall sequence is isolated, the next step is to recover the file system from any resulting modifications. The primary challenge is that attackers often exploit vulnerabilities, such as remote command execution (RCE), to create unexpected execution paths that bypass the application's standard file I/O logic. For instance, a compromised application might spawn a \texttt{curl} or \texttt{wget} process to download and write a webshell. Such actions are invisible to systems that only monitor application-level APIs.

To address this, \TITLE's file recovery module employs a two-stage strategy: \textit{active causal tracking} followed by \textit{rewind and seletive replay}. First, the \textit{Execution Dependency Analysis} stage actively constructs a provenance graph from syscalls to reveal all file operations triggered by the malicious request, directly or indirectly. Second, the \textit{Rewind \& Selective Replay} stage uses this precise attack footprint, in conjunction with backups, to surgically excise the malicious changes.

\begin{figure}[htbp]
    \centering
    \includegraphics[trim=0.6cm 0.4cm 0.6cm 0.4cm, clip,width=1\linewidth]{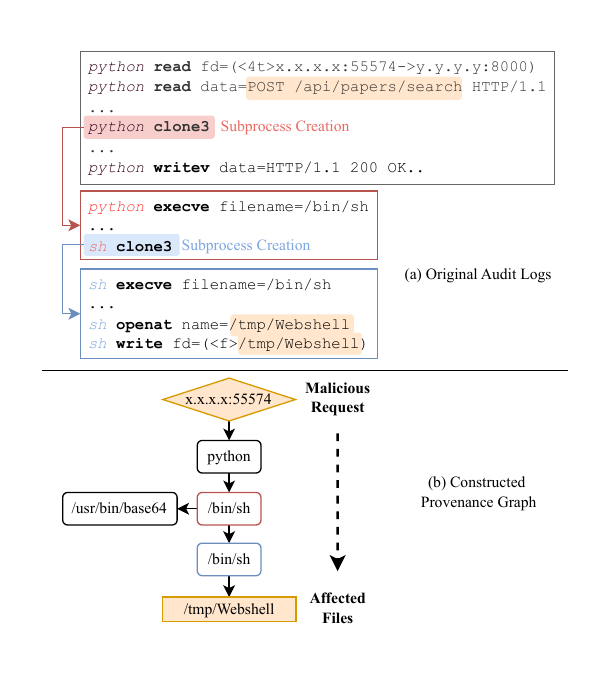}
    \caption{An example of file operation matching.}
    \label{fig:fs_example}
\end{figure}

\noindent\textbf{Execution Dependency Analysis.}
This stage constructs a complete causal chain based on syscalls to identify the full impact of a malicious request on the file system. Instead of passively monitoring predefined directories, this analysis actively constructs a provenance graph. As shown in \cref{fig:fs_example}, the analysis begins with the malicious syscall sequence from the \textit{Request Partition} module. We treat this sequence as the root of the attack and use process creation syscalls, such as \texttt{fork} and \texttt{clone}, as key events for extending the causal chain. \TITLE{} recursively tracks the activities of these newly spawned processes and threads (and their descendants) throughout the global system log.

This process builds a complete provenance graph rooted in the initial malicious request, forming a comprehensive execution chain. The graph captures all file modifications, including those performed by multi-layered descendant processes in arbitrary system locations. The output of this stage is a detailed list of affected files, each with its path, the exact operation performed, and a timestamp.

\noindent\textbf{Rewind \& Selective Replay.}
After precisely identifying all malicious file operations, the core principle of recovery is to eradicate the attack's impact while preserving all legitimate data generated concurrently. A simple system-wide rollback to a pre-attack snapshot is unacceptable as it would cause significant data loss by discarding all concurrent legitimate user operations.

\TITLE{} therefore employs a \textit{Rewind \& Selective Replay} mechanism. The process first reverts the file system to a clean state from before the attack using a snapshot, and then replays only the write operations that have been confirmed as legitimate. Since logging all file writes or backing up every file update is prohibitively expensive, we implement a differentiated recovery strategy that categorizes file system regions to reduce overhead:

\begin{itemize}[leftmargin=*]
    \item \textbf{System and Application Directories.} These areas, such as \texttt{/bin}, \texttt{/etc}, or the web application's own code directories, are expected to be static during normal operation. If the causal analysis detects that an attack has written files in these locations, \TITLE's recovery strategy is to overwrite them from a clean baseline backup.
    \item \textbf{Data Directories.} In directories for user uploads (e.g., \texttt{/uploads}) or dynamically generated content, both legitimate and malicious requests may write data. For these, \TITLE{} uses a more granular, incremental approach. It first restores an affected file to its latest pre-attack backup version. Then, it consults the log of all write operations that occurred during the attack period, filters out those identified as malicious in the previous stage, and replays the remaining legitimate writes in chronological order. This process precisely reconstructs all legitimate user data while eliminating the attack's effects. Data directories are identified either through pre-deployment analysis of the application's normal behavior or via manual configuration by an administrator.
\end{itemize}


\noindent\textbf{Implementation Details.}
     \textit{1) Provenance Graph Construction.} We use \texttt{Sysdig} as our underlying syscall collection tool, which captures detailed syscall events with low overhead. To build the provenance graph, we map these structured log events to graph components. For example, a \texttt{fork} or \texttt{clone} syscall creates a new process node and a \texttt{CREATE\_PROCESS} edge from the parent to the child. File operations like \texttt{openat} or \texttt{write} generate a file node (if not already present) and a corresponding \texttt{WRITE} or \texttt{OPEN} edge from the process node, annotated with attributes like timestamp and path. Calls like \texttt{unlink} or \texttt{rename} generate \texttt{DELETE} or \texttt{RENAME} edges, precisely capturing file state transitions. This mapping allows \TITLE{} to automatically transform the low-level syscall stream into a high-level, analyzable provenance graph.
     
     \textit{2) Write Log for Replay.} To enable granular replay for data directories, \TITLE{} captures legitimate write operations' content.
    Similar to other systems \cite{matos2021sanare}, we use a generic, application-agnostic kernel module that, upon intercepting a \texttt{write} syscall to a data directory, logs the event metadata alongside the full data payload. This logged data is buffered in-memory and written to disk asynchronously by a background thread, minimizing performance impact. These captured payloads form the ``write log'' for replay, enabling \TITLE{} to re-apply legitimate data changes onto files restored from a baseline.
     
     \textit{3) Backup and Snapshot Implementation.} \TITLE's backup strategy is twofold, designed to efficiently protect both dynamic data and static system files.
    \begin{itemize}
        \item \textit{Incremental Backup (For Data Directories):} For efficient, low-storage backups of data directories, we combine \texttt{Watchdog} (a Python library based on Linux's \texttt{inotify}) and \texttt{rsync}. \texttt{Watchdog} monitors for file changes in real-time and triggers an incremental \texttt{rsync} task. By using \texttt{rsync}'s \texttt{--link-dest} option, unchanged files in a new backup are created as hard links to their counterparts in the previous backup, rather than being copied. This significantly reduces storage consumption and enables lightweight maintenance of a file system snapshot chain.
        \item \textit{Baseline Backup (For System/Application Directories):} Static system and application directories require a ``known-good'' baseline, whose creation depends on the deployment environment. In a containerized environment, the baseline is the original, immutable container image, requiring no additional backup; \TITLE{} restores any tampered file by fetching a clean version from the read-only image layer. In a traditional VM or bare-metal environment, we perform a one-time, full backup post-initial deployment and configuration, before serving traffic. This archive is stored securely and serves as the authoritative baseline for all subsequent recoveries.
    \end{itemize}

     \textit{4) Interactive Recovery for Complex Files.} For structured files, such as binaries, where simply omitting malicious writes during replay could lead to corruption, \TITLE{} provides an interactive recovery mode. It presents the administrator with the file's last-known-good state and a detailed log of all subsequent write operations, explicitly flagging those identified as malicious. This allows the administrator to make an informed decision: perform a complete rollback, selectively replay only legitimate writes, or manually repair the file. This human-in-the-loop approach ensures the safe recovery of complex data structures, preventing file corruption that could arise from fully automated procedures.

\subsection{Post-Attack Database State Recovery}
\label{sec:db_recovery}

Restoring database state presents a unique challenge rooted in the communication patterns of modern web architectures. Web applications typically use connection pooling to communicate with databases, allowing multiple http requests to reuse the same underlying TCP connection. Concurrently, database servers employ highly asynchronous models to process requests from multiple connections. This combination of concurrency models results in a database execution stream where operations from many high-level requests are deeply interleaved. This makes it difficult to answer the critical question of \textit{which specific database operations were executed by a given malicious http request}.

To overcome this, \TITLE{} introduces a novel method for precise database operation attribution by deconstructing the asynchronous execution model. Before detailing the implementation, we first establish its underlying principle.

\begin{figure}[!t]
    \centering
    \includegraphics[trim=0.6cm 0.4cm 0.6cm 0.4cm, clip,width=1\linewidth]{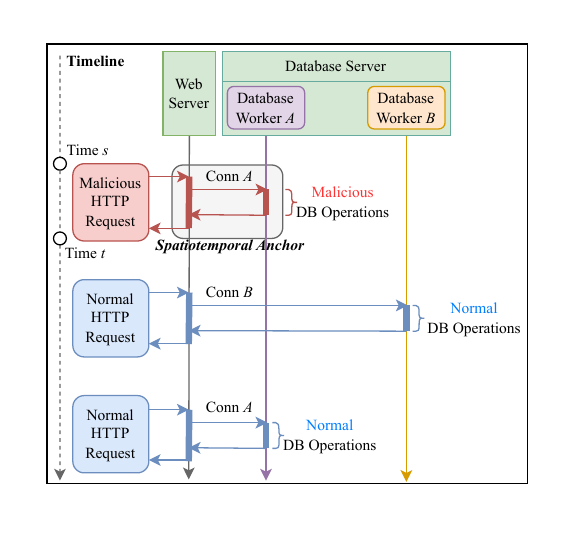}
    \caption{The principle of database operation matching.}
    \label{fig:db_match}
\end{figure}

\noindent\textbf{Principle.}
Our core insight is that by deconstructing the two layers of concurrency between a web application and its database, we can identify a stable \textit{spatiotemporal anchor} that bridges the asynchronous gap, establishing a precise causal link between a high-level request and its low-level database operations. 
We examine each concurrency model separately:
\begin{itemize}[leftmargin=*]
    \item \textit{Web Application Connection Pooling.} To avoid the high overhead of creating and destroying TCP connections, web applications use connection pools. A request ``borrows'' a connection from the pool, uses it for all its database operations, and then ``returns'' it. The key property here is \textit{exclusive use during time-division multiplexing}: although a single connection (\texttt{Conn A}) is used by different requests over its lifetime, it is used exclusively by one request for the duration of that request's processing.
    \item \textit{Database Server Concurrency.} Mainstream relational databases like \texttt{MySQL} and \texttt{PostgreSQL} typically employ a ``thread/process-per-connection'' model. When a client establishes a TCP connection, the database server dedicates a worker thread to handle all communication on that connection until it closes. This creates a stable one-to-one mapping between a network connection and a database worker thread\footnote{This model is a widely adopted architecture. We acknowledge that more complex patterns, such as the use of database proxies (e.g., ProxySQL) or advanced internal threading models in some modern databases, could disrupt this one-to-one mapping. We discuss this as a limitation in Section~\ref{sec:limitations}.}.
\end{itemize}

Combining these models reveals the path to precise attribution, as illustrated in \cref{fig:db_match}. When the \textit{Malicious Web Request} begins at \texttt{Time\_s}, it borrows \texttt{Conn\_A}. For its entire processing duration (from \texttt{Time\_s} to \texttt{Time\_t}), it has exclusive use of \texttt{Conn\_A}. On the database server, \texttt{Conn\_A} is handled exclusively by \textit{Database Worker A}. Therefore, we can conclude that within the precise time window [\texttt{Time\_s}, \texttt{Time\_t}], all database operations executed by \textit{Database Worker A} must have originated from the \textit{Malicious Web Request} and no other. This \textit{spatiotemporal anchor}, composed of the network connection tuple (identifying \texttt{Conn\_A}) and the processing time window (identifying [\texttt{Time\_s}, \texttt{Time\_t}]), is the cornerstone of our method. It allows us to deterministically attribute behavior across the two independent, asynchronous execution domains.

\smallskip

\noindent\textbf{DB Connection Analysis.}
Based on the principle above, this stage aims to extract the precise spatiotemporal anchor from the malicious request's syscall sequence. This anchor consists of the network 4-tuple (source IP, source port, destination IP, destination port) and the exact time window of its activity.

The analysis begins with the malicious syscall sequence from the \textit{Request Partition} module. \TITLE{} parses the network-related syscalls in this sequence, filtering for those targeting the database server's IP and port. Upon such identification, it extracts the full network tuple. Concurrently, it records the timestamps of the first and last syscalls using this tuple during the request's lifecycle, defining the time window. This stage outputs one or more of these anchors (\textit{Malicious Connection Fragments} in \cref{fig:overview}), serving as input for the next stage.

\smallskip

\noindent\textbf{DB Operation Extraction.}
With the spatiotemporal anchor, this stage ``fishes out'' the exact database operations triggered by the malicious request from the database server's logs. We shift our analysis to the database server's own syscall logs.

As established, a TCP connection maps to a specific database worker thread. We first use the network tuple from the anchor to identify the corresponding worker thread (via its TID) in the database server's audit logs. After locking onto the target thread, we apply the time window constraint from the anchor, examining only the syscalls executed by this thread within that window. Specifically, we focus on \texttt{write} syscalls made to the database's own application log files (e.g., \texttt{PostgreSQL}'s statement log). Since the database logs operations after executing them, the content of these \texttt{write} syscalls contains the raw text of the executed database operations. This enables \TITLE{} to precisely extract the \textit{Malicious DB Operations} from the voluminous, interleaved log I/O.

This method is broadly applicable. For some databases, such as \texttt{MongoDB}, the application-level audit log itself records the client IP and port. In such cases, the process is simplified: \TITLE{} can directly query the application log using the network and time information from the anchor, bypassing the need for syscall-level thread identification.

\smallskip

\noindent\textbf{Rewind \& Selective Replay.}
After identifying all malicious database operations, the final recovery stage aims to completely remove their effects while preserving all concurrent legitimate transactions. The process is as follows:

\begin{enumerate}[leftmargin=*]
    \item \textit{Determine Recovery Baseline.} \TITLE{} identifies the earliest timestamp among the malicious operations and selects the latest clean database snapshot (\textit{Periodic Database Backup}) created before this time.
    \item \textit{Rollback to Baseline.} The system uses the database's native tools (e.g., \texttt{pg\_restore}) to revert the entire database to the chosen clean snapshot. This erases all changes, both malicious and legitimate, that occurred after the snapshot was taken.
    \item \textit{Prepare Legitimate Operations.} To restore lost legitimate data, \TITLE{} gathers the complete set of all database operations from the database's application logs for the period between the snapshot time and the present. It then filters out the operations previously identified as malicious.
    \item \textit{Selective Replay.} Finally, \TITLE{} re-executes the remaining list of legitimate database operations, in their original chronological order, on the restored baseline database.
\end{enumerate}

Through this granular ``rollback-filter-replay'' process, \TITLE{} achieves a precise restoration of the database, producing a \textit{Recovered DB State} that is both clean and maximally preserves legitimate user data, thus avoiding the unacceptable data loss inherent in traditional snapshot-only recovery.

\subsection{System State Restoration}
\label{sec:final_state}

The final, recovered system state is achieved by composing the outcomes of the file system and database recovery processes detailed in Section~\ref{sec:file_recovery} and Section~\ref{sec:db_recovery}, respectively. Once the \textit{Rewind \& Selective Replay} mechanism has surgically repaired the file system and the database state, the system as a whole reaches the \textit{Recovered System State} shown in \cref{fig:overview}. By preserving all concurrent legitimate operations across both stateful components, \TITLE{} ensures business continuity and overcomes the critical data loss problem inherent in coarse-grained, snapshot-only rollback approaches.

\section{Evaluation}

We designed experiments to answer the following questions:

\begin{enumerate}[leftmargin=*]
\item How accurately does \TITLE match database operations and file operations to web requests under high concurrency compared to the state-of-the-art?
\item How effective is \TITLE in recovering from real-world attacks while preserving legitimate data?
\item How does \TITLE's methodology enable superior attack tracing and recovery in a practical scenario where other methods fail?
\item What is the runtime and recovery performance overhead of \TITLE?
\end{enumerate}

\subsection{Evaluation Setup}
\label{sec:evaluation_setup}
To comprehensively evaluate \TITLE, we designed experiments covering diverse modern web application ecosystems and real-world attack scenarios.

\begin{table}[htbp]
\centering
\scriptsize
\setlength{\tabcolsep}{3pt}
\caption{Vulnerability scenarios used in our experiments.}
\label{tab:attack-scenarios}
\begin{tabular}{@{}c c c c c@{}}
\toprule
\textbf{App} & \textbf{CVE ID} & \textbf{Type} & \textbf{Language} & \textbf{Database} \\
\midrule
\texttt{FastAPI} & N/A & RCE & \texttt{Python} & \texttt{PostgreSQL} \\
\midrule[0.1pt]

\multirow{2}{*}{\texttt{pgAdmin}} & CVE-2022-4223 & RCE & \multirow{2}{*}{\texttt{Python}} & \multirow{2}{*}{\texttt{PostgreSQL}} \\
                         & CVE-2023-5002 & RCE & & \\
\midrule[0.1pt]

\texttt{Django} & CVE-2019-14234 & SQLi & \texttt{Python} & \texttt{PostgreSQL} \\
\midrule[0.1pt]

\texttt{Mongo Express} & CVE-2019-10758 & RCE & \texttt{Node.js} & \texttt{MongoDB} \\
\midrule[0.1pt]

\multirow{2}{*}{\texttt{Joomla}} & CVE-2017-8917 & SQLi & \multirow{2}{*}{\texttt{PHP}} & \multirow{2}{*}{\texttt{MySQL}} \\
                        & CVE-2015-8562 & RCE & & \\
\midrule[0.1pt]

\multirow{2}{*}{\texttt{CMSMS}} & CVE-2021-26120 & RCE & \multirow{2}{*}{\texttt{PHP}} & \multirow{2}{*}{\texttt{MySQL}} \\
                       & CVE-2019-9053 & SQLi & & \\
\midrule[0.1pt]

\texttt{Gitlist} & CVE-2018-1000533 & RCE & \texttt{PHP} & No DB \\
\midrule[0.1pt]

\multirow{3}{*}{\texttt{Solr}} & CVE-2019-17558 & RCE & \multirow{3}{*}{\texttt{Java}} & \multirow{3}{*}{No DB} \\
                      & CVE-2019-0193 & RCE & & \\
                      & CVE-2017-12629-RCE & RCE & & \\
\midrule[0.1pt]

\multirow{6}{*}{\texttt{Ofbiz}} & CVE-2024-38856 & RCE & \multirow{6}{*}{\texttt{Java}} & \multirow{6}{*}{\texttt{PostgreSQL}} \\
                       & CVE-2023-51467 & RCE & & \\
                       & CVE-2023-49070 & RCE & & \\
                       & CVE-2020-9496 & RCE & & \\
                       & CVE-2024-45507 & RCE & & \\
                       & CVE-2024-45195 & RCE & & \\
\midrule[0.1pt]

\multirow{2}{*}{\texttt{GeoServer}} & CVE-2023-25157 & SQLi & \multirow{2}{*}{\texttt{Java}} & \multirow{2}{*}{\makecell{\texttt{PostGIS}\\(\texttt{PostgreSQL})}} \\
                           & CVE-2024-36401 & RCE & & \\
\bottomrule
\end{tabular}
\end{table}

\noindent\textbf{Benchmark Applications and Vulnerabilities.}
To construct a realistic evaluation environment, we selected 9 open-source web applications and one custom application (see \Cref{tab:attack-scenarios}). These applications cover four mainstream languages: \texttt{PHP}, \texttt{Python}, \texttt{Node.js}, \texttt{Java} and three database types - \texttt{MySQL}, \texttt{PostgreSQL}(both the standard version and one with the \texttt{PostGIS} extension), \texttt{MongoDB} - encompassing both SQL and NoSQL models. We reproduced 20 real-world, CVE-based attack scenarios, including high-severity vulnerabilities like SQL injection and RCE. All vulnerable environments, except our custom application, were built using images from the \cite{vulhub} project to ensure realism and reproducibility.

\noindent\textbf{Instrumentation.}
We implemented framework-level instrumentation to enable request partitioning, requiring no modification to the application code:
\begin{itemize}
\item \textit{Python}: Implemented via monkey patching, supporting \texttt{FastAPI} and other ASGI-compliant frameworks.
\item \textit{Node.js}: Implemented using the \texttt{async\_hooks} mechanism, compatible with \texttt{Express.js}.
\item \textit{Java}: Implemented using a Java Agent, compatible with thread-pool-based applications like \texttt{Solr}.
\end{itemize}

\noindent\textbf{Experimental Platform.}
All experiments were conducted on KVM-based virtual machines (Ubuntu 22.04.5, 8 vCPUs, 16GB RAM). All services were deployed in containers using Docker version 28.3.2 to ensure environmental consistency.

\noindent\textbf{Workload Generation.}
We used \texttt{Locust v2.32.6} to generate concurrent http traffic with up to 150 users, sufficient to trigger complex request interlacing. The traffic targets real APIs, ensuring that \textbf{legitimate requests also induce complex state changes in the database and file system}, creating a highly interleaved access sequence that increases the evaluation's challenge and realism.

\noindent\textbf{Ground Truth and Baselines.}
We established ground truth by running tests under single-concurrency conditions and manually labeling the causal relationships between requests and operations. We use \textsc{Sanare} \cite{matos2021sanare} as the baseline for all comparative analyses.

\subsection{Matching Accuracy}
This section quantifies \TITLE's ability to accurately correlate http requests with their underlying file and database operations under high concurrency. This precise matching is critical for surgically removing attack impacts while preserving legitimate user data. We use Precision, Recall, and F1-Score for a direct comparison against \textsc{Sanare}.

\refstepcounter{subsubsection}
\label{sec:benign-match}
\begin{table}[htbp]
\centering
\scriptsize
\caption{Precision(P), Recall(R), and F1-Score(F1) of Matching Benign Database Operations under Different Concurrency Levels: Comparison between \TITLE and \textsc{Sanare}}
\label{tab:db-match}
\renewcommand{\arraystretch}{1.2}
\setlength{\tabcolsep}{2.25pt}
\begin{tabular}{c|c|cc|cc|cc}
\midrule
\multicolumn{2}{c|}{\textbf{Concurrency}} 
& \multicolumn{2}{c|}{\textbf{1}} 
& \multicolumn{2}{c|}{\textbf{50}} 
& \multicolumn{2}{c}{\textbf{150}} \\
\midrule
\textbf{App} & \textbf{Metric} & \textsc{Ancora} & \textsc{Sanare} & \textsc{Ancora} & \textsc{Sanare} & \textsc{Ancora} & \textsc{Sanare} \\
\midrule
\multirow{3}{*}{\texttt{FastAPI}} 
& P  & 1.000 & 1.000 & \textbf{1.000} & 0.375 & \textbf{1.000} & 0.139 \\
& R  & 1.000 & 1.000 & 1.000 & 1.000 & 1.000 & 1.000 \\
& F1 & 1.000 & 1.000 & \textbf{1.000} & 0.545 & \textbf{1.000} & 0.244 \\
\hline
\multirow{3}{*}{\texttt{Django}}
& P  & 1.000 & 1.000 & \textbf{1.000} & 0.315 & \textbf{1.000} & 0.132 \\
& R  & 1.000 & 1.000 & 1.000 & 1.000 & 1.000 & 1.000 \\
& F1 & 1.000 & 1.000 & \textbf{1.000} & 0.479 & \textbf{1.000} & 0.234 \\
\hline
\multirow{3}{*}{\texttt{pgAdmin}} 
& P  & 1.000 & 1.000 & \textbf{1.000} & 0.138 & \textbf{1.000} & 0.089 \\
& R  & \textbf{1.000} & 0.722 & \textbf{1.000} & 0.630 & \textbf{1.000} & 0.621 \\
& F1 & \textbf{1.000} & 0.839 & \textbf{1.000} & 0.227 & \textbf{1.000} & 0.157 \\
\hline
\multirow{3}{*}{\texttt{Joomla}} 
& P  & 1.000 & 1.000 & \textbf{1.000} & 0.847 & \textbf{1.000} & 0.835 \\
& R  & \textbf{1.000} & 0.714 & \textbf{1.000} & 0.678 & \textbf{1.000} & 0.598 \\
& F1 & \textbf{1.000} & 0.833 & \textbf{1.000} & 0.753 & \textbf{1.000} & 0.697 \\
\hline
\multirow{3}{*}{\texttt{CMSMS}}
& P  & 1.000 & 1.000 & \textbf{1.000} & 0.352 & \textbf{1.000} & 0.382 \\
& R  & \textbf{1.000} & 0.897 & \textbf{1.000} & 0.797 & \textbf{1.000} & 0.726 \\
& F1 & \textbf{1.000} & 0.946 & \textbf{1.000} & 0.489 & \textbf{1.000} & 0.501 \\
\hline
\multirow{3}{*}{\texttt{Ofbiz}}
& P  & 1.000 & 1.000 & \textbf{1.000} & 0.225 & \textbf{1.000} & 0.224 \\
& R  & \textbf{1.000} & 0.750 & \textbf{1.000} & 0.740 & \textbf{1.000} & 0.582 \\
& F1 & \textbf{1.000} & 0.857 & \textbf{1.000} & 0.346 & \textbf{1.000} & 0.324 \\
\hline
\multirow{3}{*}{\texttt{\makecell{Mongo-\\Express}}} 
& P  & 1.000 & 1.000 & \textbf{1.000} & 0.158 & \textbf{0.987} & 0.063 \\
& R  & 1.000 & 1.000 & \textbf{1.000} & 0.666 & \textbf{0.987} & 0.829 \\
& F1 & 1.000 & 1.000 & \textbf{1.000} & 0.256 & \textbf{0.987} & 0.117 \\
\hline
\multirow{3}{*}{\texttt{GeoServer}}
& P  & 1.000 & 1.000 & \textbf{1.000} & 0.931 & \textbf{1.000} & 0.841 \\
& R  & 1.000 & 1.000 & 1.000 & 1.000 & 1.000 & 1.000 \\
& F1 & 1.000 & 1.000 & \textbf{1.000} & 0.964 & \textbf{1.000} & 0.914 \\
\midrule
\multirow{3}{*}{\textbf{Average}} 
& P  & 1.000 & 1.000 & \textbf{1.000} & 0.418 & \textbf{0.998} & 0.338 \\
& R  & \textbf{1.000} & 0.885 & \textbf{1.000} & 0.814 & \textbf{0.998} & 0.795 \\
& F1 & \textbf{1.000} & 0.936 & \textbf{1.000} & 0.507 & \textbf{0.998} & 0.399 \\
\bottomrule
\end{tabular}
\end{table}

\begin{table}[htbp]
\centering
\scriptsize
\caption{Precision(P), Recall(R), and F1-Score(F1) of Matching Benign File Operations under Different Concurrency Levels: Comparison between \TITLE and \textsc{Sanare}}
\label{tab:fs-match}
\renewcommand{\arraystretch}{1.2}
\setlength{\tabcolsep}{2.25pt}
\begin{tabular}{c|c|cc|cc|cc}
\toprule
\multicolumn{2}{c|}{\textbf{Concurrency}} 
& \multicolumn{2}{c|}{\textbf{1}} 
& \multicolumn{2}{c|}{\textbf{50}} 
& \multicolumn{2}{c}{\textbf{150}} \\
\midrule
\textbf{App} & \textbf{Metric} & \textsc{Ancora} & \textsc{Sanare} & \textsc{Ancora} & \textsc{Sanare} & \textsc{Ancora} & \textsc{Sanare} \\
\midrule
\multirow{3}{*}{\texttt{FastAPI}} 
& P  & 1.000 & 1.000 & \textbf{1.000} & 0.151 & \textbf{1.000} & 0.056 \\
& R  & 1.000 & 1.000 & 1.000 & 1.000 & 1.000 & 1.000 \\
& F1 & 1.000 & 1.000 & \textbf{1.000} & 0.262 & \textbf{1.000} & 0.106 \\
\hline
\multirow{3}{*}{\texttt{Django}}
& P  & 1.000 & 1.000 & \textbf{1.000} & 0.537 & \textbf{1.000} & 0.433 \\
& R  & 1.000 & 1.000 & 1.000 & 1.000 & \textbf{1.000} & 0.997 \\
& F1 & 1.000 & 1.000 & \textbf{1.000} & 0.699 & \textbf{1.000} & 0.604 \\
\hline
\multirow{3}{*}{\texttt{pgAdmin}} 
& P  & \textbf{1.000} & 0.963 & \textbf{1.000} & 0.520 & \textbf{1.000} & 0.804 \\
& R  & 1.000 & 1.000 & \textbf{1.000} & 0.707 & \textbf{1.000} & 0.520 \\
& F1 & \textbf{1.000} & 0.981 & \textbf{1.000} & 0.600 & \textbf{1.000} & 0.632 \\
\hline
\multirow{3}{*}{\texttt{Joomla}} 
& P  & 1.000 & 1.000 & 1.000 & 1.000 & 1.000 & 1.000 \\
& R  & 1.000 & 1.000 & \textbf{1.000} & 0.962 & \textbf{1.000} & 0.963 \\
& F1 & 1.000 & 1.000 & \textbf{1.000} & 0.980 & \textbf{1.000} & 0.981 \\
\hline
\multirow{3}{*}{\texttt{CMSMS}}
& P  & 1.000 & 1.000 & \textbf{1.000} & 0.180 & \textbf{1.000} & 0.152 \\
& R  & 1.000 & 1.000 & 1.000 & 1.000 & 1.000 & 1.000 \\
& F1 & 1.000 & 1.000 & \textbf{1.000} & 0.304 & \textbf{1.000} & 0.264 \\
\hline
\multirow{3}{*}{\texttt{Gitlist}}
& P  & 1.000 & 1.000 & \textbf{1.000} & 0.064 & \textbf{1.000} & 0.029 \\
& R  & 1.000 & 1.000 & 1.000 & 1.000 & 1.000 & 1.000 \\
& F1 & 1.000 & 1.000 & \textbf{1.000} & 0.121 & \textbf{1.000} & 0.056 \\
\hline
\multirow{3}{*}{\texttt{Solr}} 
& P  & \textbf{1.000} & 0.000 & \textbf{1.000} & 0.000 & \textbf{1.000} & 0.000 \\
& R  & \textbf{1.000} & 0.000 & \textbf{1.000} & 0.000 & \textbf{1.000} & 0.000 \\
& F1 & \textbf{1.000} & 0.000 & \textbf{1.000} & 0.000 & \textbf{1.000} & 0.000 \\
\hline
\multirow{3}{*}{\texttt{Ofbiz}}
& P  & 1.000 & 1.000 & \textbf{1.000} & 0.027 & \textbf{1.000} & 0.014 \\
& R  & 1.000 & 1.000 & \textbf{1.000} & 0.824 & \textbf{1.000} & 0.726 \\
& F1 & 1.000 & 1.000 & \textbf{1.000} & 0.052 & \textbf{1.000} & 0.027 \\
\hline
\multirow{3}{*}{\texttt{GeoServer}}
& P  & 1.000 & 1.000 & \textbf{1.000} & 0.214 & \textbf{1.000} & 0.091 \\
& R  & \textbf{1.000} & 0.586 & \textbf{1.000} & 0.455 & \textbf{1.000} & 0.512 \\
& F1 & \textbf{1.000} & 0.739 & \textbf{1.000} & 0.291 & \textbf{1.000} & 0.154 \\
\midrule
\multirow{3}{*}{\textbf{Average}} 
& P  & \textbf{1.000} & 0.885 & \textbf{1.000} & 0.299 & \textbf{1.000} & 0.287 \\
& R  & \textbf{1.000} & 0.843 & \textbf{1.000} & 0.772 & \textbf{1.000} & 0.858 \\
& F1 & \textbf{1.000} & 0.858 & \textbf{1.000} & 0.368 & \textbf{1.000} & 0.314 \\
\bottomrule
\end{tabular}
\end{table}

\noindent 1) \textbf{Matching Accuracy for Benign Operations.}

As shown in \Cref{tab:db-match} and \Cref{tab:fs-match}, \TITLE{} demonstrates superior accuracy and stability when processing benign requests.
\begin{itemize}
\item \textbf{\TITLE{}}: Achieved nearly perfect matching accuracy (F1-Score=1.0) across almost all tests. A minor dip in \texttt{Mongo-Express} (F1-Score of 0.987) was caused by occasional event loss in the underlying \texttt{sysdig} tool under extreme load, not a flaw in \TITLE's mechanism.
\item \textbf{\textsc{Sanare}}: Accuracy deteriorated sharply with increasing concurrency. As concurrency rose from 1 to 150, its F1-score for file and database operation matching dropped by an average of 68.63\% and 60.25\%, respectively. For instance, in \texttt{FastAPI}'s database matching, \textsc{Sanare}'s precision plummeted from 100\% to 13.91\%.
\end{itemize}
These results show that \textsc{Sanare}'s black-box approach fails to learn stable patterns in ``noisy'' high-concurrency environments, leading to massive misattributions. In contrast, \TITLE's deterministic correlation method is immune to such interference.

\smallskip
\noindent 2) \textbf{Matching Accuracy for Vulnerability-Exploitation Operations.}

\TITLE's advantage is even more pronounced in real-world attack scenarios, especially those involving RCE.

\begin{table}[htbp]
\centering
\scriptsize
\caption{F1-Score of Matching Malicious Operations under Different Concurrency Levels: Comparison between \textsc{Ancora} and \textsc{Sanare}}
\label{tab:malicious-match}
\renewcommand{\arraystretch}{1.2}
\setlength{\tabcolsep}{0.8pt}
\begin{tabular}{c|cc|cc|cc}
\hline
{\textbf{Concurrency}} 
& \multicolumn{2}{c|}{\textbf{1}} 
& \multicolumn{2}{c|}{\textbf{50}} 
& \multicolumn{2}{c}{\textbf{150}} \\
\hline
\textbf{App (Vulnerability)} & \textsc{Ancora} & \textsc{Sanare} & \textsc{Ancora} & \textsc{Sanare} & \textsc{Ancora} & \textsc{Sanare} \\
\hline
{\makecell{Self-Design, RCE}} 
& \textbf{1.000} & 0.000 & \textbf{1.000} & 0.000 & \textbf{0.999} & 0.000 \\
\hline
{\makecell{CVE-2022-4223, RCE}} 
& \textbf{1.000} & 0.000 & \textbf{1.000} & 0.000 & \textbf{1.000} & 0.000 \\
\hline
{\makecell{CVE-2023-5002, RCE}} 
& \textbf{1.000} & 0.000 & \textbf{1.000} & 0.000 & \textbf{1.000} & 0.000 \\
\hline
{\makecell{CVE-2015-8562, RCE}} 
& \textbf{1.000} & 0.000 & \textbf{1.000} & 0.000 & \textbf{0.993} & 0.000 \\
\hline
{\makecell{CVE-2021-26120, RCE}}
& \textbf{1.000} & 0.000 & \textbf{1.000} & 0.000 & \textbf{1.000} & 0.000 \\
\hline
{\makecell{CVE-2018-1000533, RCE}}
& \textbf{1.000} & 0.000 & \textbf{1.000} & 0.000 & \textbf{1.000} & 0.000 \\
\hline
{\makecell{CVE-2019-17558, RCE}}
& \textbf{1.000} & 0.000 & \textbf{1.000} & 0.000 & \textbf{1.000} & 0.000 \\
\hline
{\makecell{CVE-2019-0193, RCE}}
& \textbf{1.000} & 0.000 & \textbf{1.000} & 0.000 & \textbf{1.000} & 0.000 \\
\hline
{\makecell{CVE-2017-12629, RCE}}
& \textbf{1.000} & 0.000 & \textbf{1.000} & 0.000 & \textbf{0.994} & 0.000 \\
\hline
{\makecell{CVE-2024-38856, RCE}}
& \textbf{1.000} & 0.000 & \textbf{1.000} & 0.000 & \textbf{1.000} & 0.000 \\
\hline
{\makecell{CVE-2023-51467, RCE}}
& \textbf{1.000} & 0.000 & \textbf{1.000} & 0.000 & \textbf{1.000} & 0.000 \\
\hline
{\makecell{CVE-2023-49070, RCE}}
& \textbf{1.000} & 0.000 & \textbf{1.000} & 0.000 & \textbf{1.000} & 0.000 \\
\hline
{\makecell{CVE-2020-9496, RCE}}
& \textbf{1.000} & 0.000 & \textbf{1.000} & 0.000 & \textbf{1.000} & 0.000 \\
\hline
{\makecell{CVE-2024-45507, RCE}}
& \textbf{1.000} & 0.000 & \textbf{1.000} & 0.000 & \textbf{1.000} & 0.000 \\
\hline
{\makecell{CVE-2024-45195, RCE}}
& \textbf{1.000} & 0.000 & \textbf{1.000} & 0.000 & \textbf{1.000} & 0.000 \\
\hline
{\makecell{CVE-2024-36401, RCE}}
& \textbf{1.000} & 0.000 & \textbf{1.000} & 0.000 & \textbf{1.000} & 0.000 \\
\hline
{\makecell{CVE-2023-25157, SQLi}}
& \textbf{1.000} & 0.000 & \textbf{1.000} & 0.000 & \textbf{1.000} & 0.000 \\
\hline
{\makecell{CVE-2019-14234, SQLi}}
& \textbf{1.000} & 0.000 & \textbf{1.000} & 0.000 & \textbf{0.978} & 0.000 \\
\hline
{\textbf{Average}} 
& \textbf{1.000} & 0.000 & \textbf{1.000} & 0.000 & \textbf{0.998} & 0.000 \\
\hline
\end{tabular}
\end{table}

\begin{itemize}
\item \textbf{\TITLE{}}: As shown in \Cref{tab:malicious-match}, \TITLE{} maintained a near-perfect F1-Score of 1.0 across all 17 real-world vulnerability scenarios \footnote{\textsc{Sanare} could not be trained on \texttt{Mongo Express} due to the absence of normal file operation data in the experimental scenario, and the two SQL injection vulnerabilities CVE-2019-9053 and CVE-2017-8917 could not be exploited to inject statements that modify the database.} and all concurrency levels. This confirms its deterministic tracking can precisely follow unexpected execution paths created by exploits.
\item \textbf{\textsc{Sanare}}: Completely failed to handle these attacks, with its F1-Score remaining at 0 in all tests. Its machine learning model could not recognize malicious inputs absent from training data, nor could it track execution paths that deviated from normal logic.
\end{itemize}
These experiments clearly demonstrate that \TITLE's matching method maintains near-perfect accuracy against realistic exploits, whereas the black-box approach fails entirely due to its inherent design limitations. This highlights \TITLE's decisive advantage in handling complex attack scenarios.

\subsection{Recovery Effectiveness}
This experiment evaluates the final recovery outcome. We use a strict accuracy criterion: a recovery is accurate only if the set of restored operations $\mathcal{P}$ is exactly equal to the ground-truth set $\mathcal{Q}$ (i.e., $\mathcal{P}=\mathcal{Q}$), ensuring no omissions or collateral damage. The accuracy is defined as:
\begin{equation}
\label{eq:accuracy}
\textit{Accuracy}=\frac{\textit{\# of accurately recovered requests}}{\textit{\# of total requests}}.
\end{equation}

\begin{table*}[htbp]
\centering
\scriptsize
\caption{Database Operations Recovery Accuracy (\%) of \textsc{Ancora} and \textsc{Sanare} under Different Concurrency Levels}
\label{tab:db-recovery-accuracy}
\renewcommand{\arraystretch}{1.2}
\setlength{\tabcolsep}{4pt}
\begin{tabular}{c|cc|cc|cc|cc|cc|cc}
\toprule
{\textbf{Concurrency}} 
& \multicolumn{2}{c|}{\textbf{1}} 
& \multicolumn{2}{c|}{\textbf{5}} 
& \multicolumn{2}{c|}{\textbf{10}} 
& \multicolumn{2}{c|}{\textbf{50}} 
& \multicolumn{2}{c|}{\textbf{100}} 
& \multicolumn{2}{c}{\textbf{150}} \\
\midrule
\textbf{APP}
& \textsc{Ancora} & \textsc{Sanare} 
& \textsc{Ancora} & \textsc{Sanare} 
& \textsc{Ancora} & \textsc{Sanare} 
& \textsc{Ancora} & \textsc{Sanare} 
& \textsc{Ancora} & \textsc{Sanare} 
& \textsc{Ancora} & \textsc{Sanare} \\
\midrule
\texttt{FastAPI} 
& 100.00 & 100.00 & \textbf{100.00} & 34.48 & \textbf{100.00} & 28.67 & \textbf{100.00} & 36.27 & \textbf{100.00} & 26.00 & \textbf{100.00} & 33.65 \\
\texttt{Django} 
& 100.00 & 100.00 & \textbf{100.00} & 86.73 & \textbf{100.00} & 75.53 & \textbf{100.00} & 57.40 & \textbf{100.00} & 42.80 & \textbf{99.79} & 39.72 \\
\texttt{pgAdmin} 
& \textbf{100.00} & 72.22 & \textbf{100.00} & 48.31 & \textbf{100.00} & 36.78 & \textbf{100.00} & 8.10 & \textbf{100.00} & 5.86 & \textbf{100.00} & 5.50 \\
\texttt{Joomla} 
& \textbf{100.00} & 50.00 & \textbf{100.00} & 30.77 & \textbf{100.00} & 25.88 & \textbf{100.00} & 26.43 & \textbf{100.00} & 22.73 & \textbf{100.00} & 19.62 \\
\texttt{CMSMS} 
& \textbf{100.00} & 72.73 & \textbf{100.00} & 7.14 & \textbf{100.00} & 0.98 & \textbf{100.00} & 0.00 & \textbf{100.00} & 1.45 & \textbf{100.00} & 0.83 \\
\texttt{Ofbiz} 
& \textbf{100.00} & 0.00 & \textbf{100.00} & 0.00 & \textbf{100.00} & 0.00 & \textbf{100.00} & 0.00 & \textbf{100.00} & 0.00 & \textbf{100.00} & 0.00 \\
\texttt{Mongo Express} 
& 100.00 & 100.00 & \textbf{100.00} & 47.22 & \textbf{100.00} & 29.89 & \textbf{100.00} & 10.92 & \textbf{98.73} & 10.38 & \textbf{98.70} & 9.73 \\
\texttt{GeoServer} 
& 100.00 & 100.00 & 100.00 & 100.00 & \textbf{100.00} & 98.97 & \textbf{100.00} & 93.14 & \textbf{100.00} & 84.87 & \textbf{100.00} & 84.35 \\
\midrule
\textbf{Average} 
& \textbf{100.00} & 74.37 & \textbf{100.00} & 44.33 & \textbf{100.00} & 37.09 
& \textbf{100.00} & 29.03 & \textbf{99.84} & 24.26 & \textbf{99.81} & 24.18 \\
\bottomrule
\end{tabular}
\end{table*}

\begin{table*}[htbp]
\centering
\scriptsize
\caption{File Operations Recovery Accuracy (\%) of \textsc{Ancora} and \textsc{Sanare} under Different Concurrency Levels}
\label{tab:file-recovery-accuracy}
\renewcommand{\arraystretch}{1.2}
\setlength{\tabcolsep}{4pt}
\begin{tabular}{c|cc|cc|cc|cc|cc|cc}
\toprule
{\textbf{Concurrency}} 
& \multicolumn{2}{c|}{\textbf{1}} 
& \multicolumn{2}{c|}{\textbf{5}} 
& \multicolumn{2}{c|}{\textbf{10}} 
& \multicolumn{2}{c|}{\textbf{50}} 
& \multicolumn{2}{c|}{\textbf{100}} 
& \multicolumn{2}{c}{\textbf{150}} \\
\midrule
\textbf{APP}
& \textsc{Ancora} & \textsc{Sanare} 
& \textsc{Ancora} & \textsc{Sanare} 
& \textsc{Ancora} & \textsc{Sanare} 
& \textsc{Ancora} & \textsc{Sanare} 
& \textsc{Ancora} & \textsc{Sanare} 
& \textsc{Ancora} & \textsc{Sanare} \\
\midrule
\texttt{FastAPI} 
& 100.00 & 100.00 & \textbf{100.00} & 66.00 & \textbf{100.00} & 44.83 & \textbf{100.00} & 43.48 & \textbf{100.00} & 39.46 & \textbf{100.00} & 37.44 \\
\texttt{Django} 
& 100.00 & 100.00 & \textbf{100.00} & 90.83 & \textbf{100.00} & 86.67 & \textbf{100.00} & 67.97 & \textbf{100.00} & 67.72 & \textbf{100.00} & 62.57 \\
\texttt{pgAdmin} 
& \textbf{100.00} & 94.44 & \textbf{100.00} & 73.26 & \textbf{100.00} & 41.28 & \textbf{100.00} & 36.45 & \textbf{100.00} & 44.09 & \textbf{100.00} & 37.90 \\
\texttt{Joomla} 
& 100.00 & 100.00 & 100.00 & 100.00 & \textbf{100.00} & 98.33 & \textbf{100.00} & 95.23 & \textbf{100.00} & 93.59 & \textbf{100.00} & 95.61 \\
\texttt{CMSMS} 
& 100.00 & 100.00 & \textbf{100.00} & 46.27 & \textbf{100.00} & 25.69 & \textbf{100.00} & 1.38 & \textbf{100.00} & 1.96 & \textbf{100.00} & 1.18 \\
\texttt{Gitlist} 
& 100.00 & 100.00 & \textbf{100.00} & 22.40 & \textbf{100.00} & 3.47 & \textbf{100.00} & 0.00 & \textbf{100.00} & 0.00 & \textbf{100.00} & 0.00 \\
\texttt{Solr} 
& \textbf{100.00} & 0.00 & \textbf{100.00} & 0.00 & \textbf{100.00} & 0.00 & \textbf{100.00} & 0.00 & \textbf{100.00} & 0.00 & \textbf{100.00} & 0.00 \\
\texttt{Ofbiz} 
& 100.00 & 100.00 & \textbf{100.00} & 4.00 & \textbf{100.00} & 0.67 & \textbf{100.00} & 0.00 & \textbf{100.00} & 0.00 & \textbf{100.00} & 0.00 \\
\texttt{GeoServer} 
& \textbf{100.00} & 0.00 & \textbf{100.00} & 0.00 & \textbf{100.00} & 0.00 & \textbf{100.00} & 0.00 & \textbf{100.00} & 0.00 & \textbf{100.00} & 0.00 \\
\midrule
\textbf{Average} 
& \textbf{100.00} & 77.16 & \textbf{100.00} & 44.31 & \textbf{100.00} & 33.55 
& \textbf{100.00} & 27.39 & \textbf{100.00} & 27.31 & \textbf{100.00} & 26.08 \\
\bottomrule
\end{tabular}
\end{table*}

As summarized in \Cref{tab:db-recovery-accuracy} and \Cref{tab:file-recovery-accuracy}, \TITLE's recovery effectiveness far surpasses the baseline.
\begin{itemize}
\item \textbf{\TITLE{}}: Attained 100\% recovery accuracy across the vast majority of applications and concurrency levels, validating its robustness. Minor dips (e.g., 99.79\% for \texttt{Django} at 150 concurrency) are consistent with occasional event loss from the underlying logging tool.
\item \textbf{\textsc{Sanare}}: Performance degraded drastically as concurrency increased. Its average recovery accuracy for file operations dropped from 77.16\% at single concurrency to 26.08\% at 150 concurrency; for database operations, the average fell from 74.37\% to 24.18\%. In worst-case scenarios like \texttt{Gitlist}, accuracy fell to 0\% at just 50 concurrency.
\end{itemize}

This performance gap stems from the systems' fundamental principles. In high-concurrency environments, operations from different requests are tightly interleaved. \textsc{Sanare}'s statistical model is susceptible to this ``crosstalk'', erroneously associating temporally adjacent operations and thus failing our strict criterion. In contrast, \TITLE's deterministic causal tracing, built on provenance graphs, creates isolated causal chains for each request. This guarantees the atomicity and accuracy of recovery, maintaining exceptionally high fidelity even in highly concurrent scenarios.

\subsection{Case Study}
We present a case study on a multi-stage attack on \texttt{CMS Made Simple v2.2.9.1} to evaluate \TITLE's accuracy in a realistic scenario. The attack chain leverages CVE-2021-26120 to achieve remote code execution (RCE), causing modifications to both the database and the file system.

\noindent 1) \textbf{Attack Chain.}
The attack, requiring administrator privileges, proceeds in two stages:
\begin{itemize}
\item \textbf{Stage 1 (Template Injection).} The attacker injects a Server-Side Template Injection (SSTI) payload into a template via the backend editor. The payload exploits a vulnerability in the Smarty engine:
\begin{lstlisting}[
  basicstyle=\ttfamily\footnotesize, breaklines=true, frame=single,
  xleftmargin=1em, framexleftmargin=0.5em, columns=flexible,
  keepspaces=true, language={},
]
{function name='rce(){}; 
    system("curl -o /tmp/Webshell http://..."); 
function '}
{/function}endrce
\end{lstlisting}

\item\textbf{Stage 2 (RCE and Backdoor).} When a user visits the page rendering this template, the server executes the payload, which downloads and saves a webshell.
\end{itemize}

This attack overwrites template content in the database and creates a malicious file on the file system.

\smallskip

\noindent 2) \textbf{Recovery Evaluation.}
We labeled the two attack requests as malicious and triggered recovery for both \textsc{Sanare} and \TITLE{}. To create a high-noise environment, we simulated 30 concurrent users for one minute, generating heavy, interleaved background traffic (e.g., page visits, configuration updates). This workload produced 385 total http requests, 374.43 MiB of syscall logs (598,432 events), 5,398 database write operations, and 2,128 file system modification operations. The recovery goal was to precisely revert the two malicious database writes and the single webshell file creation amidst this activity.

\smallskip
We first evaluated \textsc{Sanare}, which failed in both stages. In the first stage (database recovery), the attack caused only two database writes. However, \textsc{Sanare}'s time-window-based approach incorrectly identified 13 additional, legitimate writes from background traffic as malicious. This resulted in an \textbf{87.5\% false positive rate} (13 out of 15), causing irreversible state corruption by rolling back valid user data. In the second stage (file system recovery), \textsc{Sanare} completely missed the webshell creation. The execution path (RCE leading to a file write) was not present in its training data, so it could not associate the request with the file operation. Consequently, the webshell persisted after the recovery attempt.

\begin{figure}[!t]
    \centering
    \includegraphics[trim=0.6cm 0.4cm 0.6cm 0.6cm, clip,width=0.85\linewidth]{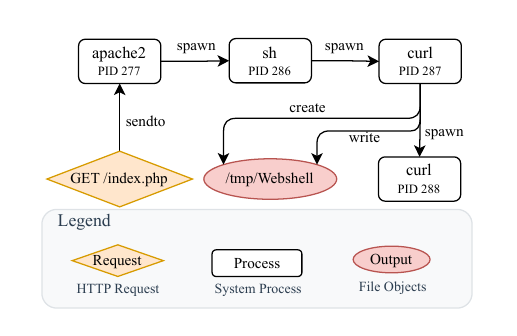}
    \caption{File Provenance Graph of Case Study. By tracing the causal chain of process spawning and file I/O operations, \TITLE precisely links the initial malicious http request to the creation of the hidden webshell.}
    \label{fig:provenance}
\end{figure}

In contrast, \TITLE \textbf{achieved precise recovery in both stages}. For the database recovery, \TITLE{} correctly identified only the two malicious writes by tracking the request to its specific database connection (\texttt{172.18.0.3:50564 -> 172.18.0.2:3306}) and worker thread (\texttt{ID=337278}), precisely isolating the operations within their execution window (\texttt{T\_start=<15:00:19.700041883>, T\_end=<15:00:19.810219831>}). No legitimate data was affected. Similarly, for the file system recovery, \TITLE{} correctly identified the webshell creation. It constructed a full provenance graph from syscalls (\Cref{fig:provenance}) that linked the http request through the \texttt{PHP} and forked \texttt{curl} processes to the final file write, allowing the recovery process to correctly delete the webshell without affecting other files.

In summary, this case study demonstrates that \TITLE{} can accurately recover from a multi-stage attack with interleaved background traffic. While \textsc{Sanare} suffered from a high false positive rate in database recovery and completely missed the file system damage, \TITLE{}'s fine-grained causality tracking enabled a complete and precise restoration of the system state.

\subsection{Performance Overhead}

We evaluate the performance overhead of \TITLE{} in two phases: runtime and recovery. All experiments are conducted in the environment described in \Cref{sec:evaluation_setup}.

\noindent 1) \textbf{Runtime Overhead.}
Runtime overhead stems from two sources: computation for request attribution and recovery, and I/O for state backup.

\smallskip

\noindent \textit{a) Computation Overhead.}

The computational overhead is caused by three main activities: (1) framework instrumentation for request partitioning (e.g., monkey patching \texttt{FastAPI}); (2) real-time syscall tracing with \texttt{sysdig} for causality tracking; and (3) periodic backup of the file system and database.

We stress-tested \TITLE{} using \texttt{Locust} with concurrency up to 150, targeting API endpoints that trigger database and file system writes. In the worst-case scenario, \TITLE{} introduced a 19.8\% increase in average response latency and a 17.8\% decrease in Queries Per Second (QPS). As shown in \Cref{fig:qps_and_response_time}, the overhead varies with application type and concurrency:

\begin{figure}[!t]
    \centering
    \includegraphics[trim=0.2cm 0.2cm 0.2cm 0.2cm, clip,width=0.9\linewidth]{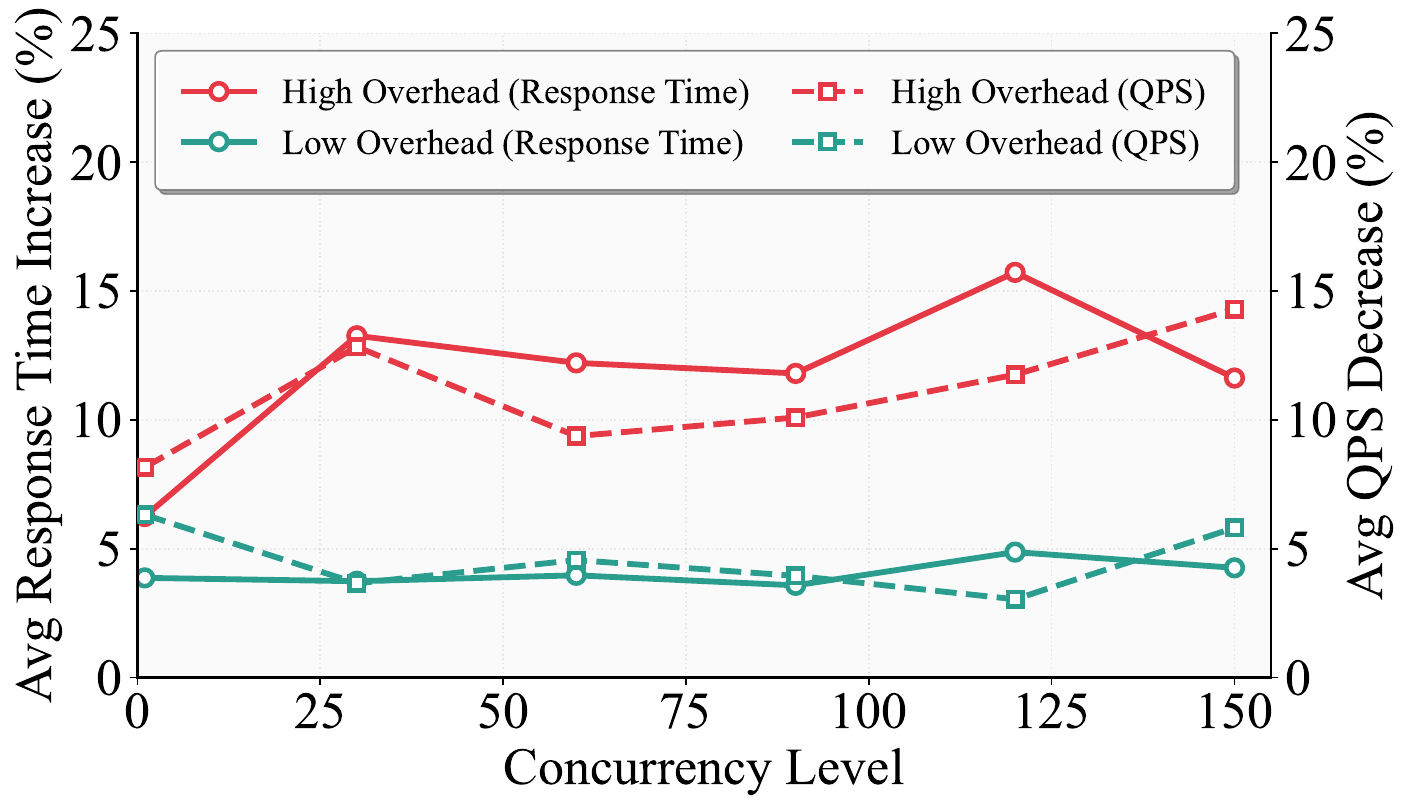}
    \caption{Avg Response Time Increase and QPS Decrease for High-Overhead (\texttt{CMSMS}, \texttt{pgAdmin}, \texttt{Django}) and Low-Overhead Applications under Increasing Concurrency Levels.}
    \label{fig:qps_and_response_time}
\end{figure}

\begin{itemize}
\item \textbf{High-overhead Applications:} For I/O-intensive or complex applications like \texttt{CMSMS}, \texttt{Django}, and \texttt{pgAdmin}, the response time increased by 6.25\%--15.72\%, and QPS decreased by 8.16\%--14.29\% as concurrency increased.

\item \textbf{Low-overhead Applications:} For the remaining 7 applications, the overhead was minimal and stable. The average response time increase was 3.5\%--4.9\%, and the QPS decrease was 3.0\%--6.3\% at all concurrency levels.
\end{itemize}

In summary, the performance overhead of \TITLE remains within a reasonable range, demonstrating that its design achieves precise recovery capabilities while maintaining operational efficiency.

\smallskip

\noindent \textit{b) Storage and I/O Overhead.}

\begin{table*}[htbp]
\centering
\scriptsize
\caption{Per-request Storage Overhead (KiB) for Different Applications under Various Conditions}
\label{tab:storage-overhead}
\renewcommand{\arraystretch}{1.2}
\setlength{\tabcolsep}{4pt}
\begin{tabular}{c|c|c|c|c|c|c|c|c|c|c}
\toprule
\textbf{App} 
& \texttt{GeoServer} & \texttt{Ofbiz} & \texttt{Solr} & \texttt{GitList} & \texttt{CMSMS} 
& \texttt{Joomla} & \texttt{Django} & \texttt{pgAdmin} & \texttt{FastAPI} & \texttt{Mongo Express} \\
\midrule
\textbf{Storage Overhead (KiB)} 
& 118.77 & 1676.50 & 460.40 & 613.12 & 1615.27 & 612.43 & 94.81 & 190.74 & 344.71 & 78.07 \\
\bottomrule
\end{tabular}
\end{table*}
This overhead is induced by \texttt{sysdig} logging, state backups, and our write-log kernel module. We measured storage consumption across tens of thousands of mixed requests on the ten applications from \Cref{tab:attack-scenarios}. On average, \TITLE{} consumed 580.48 KiB of storage per request, with a maximum of 1.64 MiB. The per-request overhead, detailed in \Cref{tab:storage-overhead}, varies significantly with application I/O patterns:
\begin{itemize}
    \item \textbf{Lightweight:} \texttt{Django} and \texttt{Mongo Express} required only 94.81 KiB and 78.07 KiB per request, respectively.
    \item \textbf{Moderate:} Most applications, such as \texttt{GeoServer} (118.77 KiB), \texttt{pgAdmin} (190.74 KiB), and \texttt{Solr} (460.40 KiB), fell into a medium range.
    \item \textbf{I/O-intensive:} Complex applications like \texttt{Ofbiz} and \texttt{CMSMS} incurred the highest overhead, at 1.64 MiB and 1.58 MiB per request, respectively.
\end{itemize}

\noindent 2) \textbf{Recovery Overhead.}
We evaluate the efficiency and scalability of \TITLE's ``rollback–filter–replay'' recovery mechanism. To simulate attacks, we label a subset of requests as ``malicious'' to trigger recovery and measure the \textbf{recovery time} and \textbf{scalability}.

\smallskip

\noindent \textit{a) Recovery Time.}

Recovery time is the duration to roll back the system state and replay legitimate operations. This time depends on the application and the extent of the damage.

\begin{table}[htbp]
\centering
\scriptsize
\caption{Database Operations Recovery Time (ms) under Different Operation Counts}
\label{tab:db-recovery-time}
\renewcommand{\arraystretch}{1.2}
\setlength{\tabcolsep}{3pt}
\begin{tabular}{c|c|c|c|c|c}
\toprule
\textbf{App} 
& \textbf{200} 
& \textbf{400} 
& \textbf{600} 
& \textbf{800} 
& \textbf{1000} \\
\midrule
\texttt{Mongo Express} 
& 2233.142 & 3541.279 & 5016.447 & 6663.838 & 7981.847 \\
\texttt{FastAPI} 
& 881.530 & 1112.621 & 1334.913 & 1570.776 & 1825.448 \\
\texttt{Django} 
& 597.667 & 746.039 & 876.747 & 997.427 & 1167.025 \\
\texttt{pgAdmin} 
& 913.301 & 1064.009 & 1211.451 & 1478.809 & 1615.741 \\
\texttt{Joomla} 
& 1108.840 & 1254.408 & 1336.578 & 1458.148 & 1560.833 \\
\texttt{CMSMS} 
& 2828.231 & 3401.275 & 4064.171 & 5076.487 & 5581.244 \\
\texttt{Ofbiz} 
& 44248.147 & 44832.673 & 45158.676 & 45745.375 & 46209.707 \\
\texttt{GeoServer} 
& 5218.507 & 5415.005 & 5536.577 & 5811.295 & 6276.799 \\
\midrule
\textbf{Average} 
& 7252.421 & 7670.925 & 8066.945 & 8600.269 & 9027.331 \\
\bottomrule
\end{tabular}
\end{table}

\begin{table}[htbp]
\centering
\scriptsize
\caption{File System Recovery Time (ms) under Different Numbers of Affected Files}
\label{tab:fs-recovery-time}
\renewcommand{\arraystretch}{1.2}
\setlength{\tabcolsep}{4.5pt}
\begin{tabular}{c|c|c|c|c|c}
\toprule
\textbf{App} 
& \textbf{20} 
& \textbf{40} 
& \textbf{60} 
& \textbf{80} 
& \textbf{100} \\
\midrule
\texttt{FastAPI} 
& 37.654 & 62.901 & 92.632 & 116.481 & 145.346 \\
\texttt{Django} 
& 122.873 & 237.509 & 345.362 & 479.495 & 580.510 \\
\texttt{pgAdmin} 
& 106.504 & 223.449 & 328.780 & 412.827 & 539.478 \\
\texttt{Joomla} 
& 828.145 & 1754.703 & 2495.804 & 3400.287 & 4047.482 \\
\texttt{CMSMS} 
& 1547.066 & 2676.764 & 4241.084 & 5308.870 & 6765.862 \\
\texttt{Gitlist} 
& 237.192 & 458.865 & 664.263 & 862.598 & 1094.465 \\
\texttt{Solr} 
& 3248.172 & 6816.581 & 8415.476 & 11592.727 & 13923.036 \\
\texttt{GeoServer} 
& 747.417 & 1084.144 & 1555.632 & 1889.531 & 2261.270 \\
\midrule
\textbf{Average} 
& 859.378 & 1664.365 & 2267.379 & 3007.852 & 3669.681 \\
\bottomrule
\end{tabular}
\end{table}

As shown in \Cref{tab:db-recovery-time,tab:fs-recovery-time}, recovering from 1,000 malicious database operations took 9.03 s on average (max 46.2 s). Restoring 100 affected files took 3.67 s on average (max 13.9 s). 
\begin{itemize}
    \item \textbf{Database recovery:} Replaying 1,000 legitimate operations was fast for lightweight applications like \texttt{Django} (1.17 s) and \texttt{Joomla} (1.56 s), but took 46.2 s for the transaction-heavy \texttt{Ofbiz}.
    \item \textbf{File system recovery:} Restoring 100 files took 1.1 s in \texttt{Gitlist} (simple creation/deletion) but 13.9 s in \texttt{Solr} (complex modifications).
\end{itemize}

\smallskip

\noindent \textit{b) Scalability of Recovery.}
\begin{figure}[!t]
    \centering
    \includegraphics[trim=0.2cm 0.2cm 0.2cm 0.2cm, clip,width=1\linewidth]{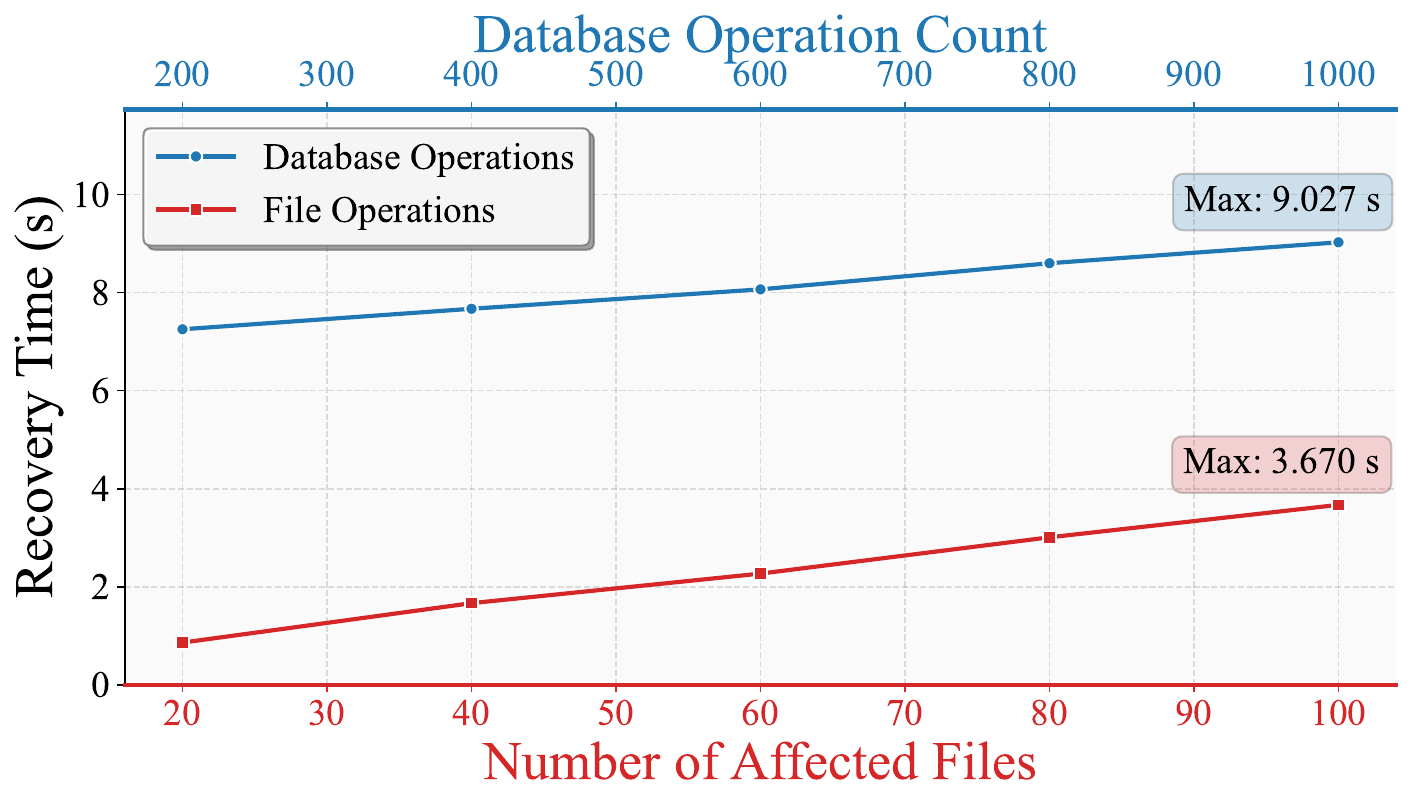}
    \caption{Recovery Time Scaling with Increasing Workload}
    \label{fig:recovery_time}
\end{figure}

We evaluated scalability by measuring recovery time against an increasing number of entities to restore. As shown in \cref{fig:recovery_time}, \TITLE{} exhibits excellent scalability. The recovery time for both database and file system restoration scales linearly with the number of replayed operations and repaired files. This demonstrates that the recovery overhead is predictable and manageable even for large-scale incidents.

\section{Limitation and Discussion}
\label{sec:limitations}

1) \textbf{Advanced Database Architectures.} 
\TITLE's database operation tracing relies on a stable, one-to-one mapping between a client network connection and a database worker thread. While effective for standard deployments of \texttt{MySQL} and \texttt{PostgreSQL}, this assumption is violated by architectures involving database proxies (e.g., \texttt{ProxySQL}) or advanced concurrency models. These intermediaries obscure the client connection, breaking our attribution method. Extending \TITLE{} to these settings is a direction for future work.

2) \textbf{Recovery of Remote Services.}
\TITLE{} is designed to restore the server's local state (file system and database). It can identify malicious outbound requests to external web services (e.g., payment gateways) but cannot automatically revert their effects. While \TITLE{} provides the forensic data for manual compensation, developing frameworks for automated cross-system remediation remains an open challenge.

3) \textbf{Coverage of Other Internal State.} 
Our current model focuses on file systems and databases. It does not cover other forms of state, such as distributed caches (e.g., \texttt{Redis}) or message queues (e.g., \texttt{Kafka}). An attack could poison these components, leading to delayed or cross-service damage. Extending \TITLE's causal tracking and recovery mechanisms to stateful middlewares is an avenue for future research.

\section{Related Work}
\label{sec:related_work}

Research on web application intrusion recovery can be broadly categorized.

\begin{itemize}[leftmargin=*]
\item\textbf{Taint-based Analysis.} 
Systems like \textsc{Shuttle} \cite{nascimento2015shuttle} use deep instrumentation of the application runtime to propagate taint markers from http requests to low-level operations. This approach, while precise in theory, is highly invasive, making it difficult to deploy and maintain in complex production applications. It also fails to track operations from unexpected execution paths created by vulnerabilities like RCE. In contrast, \TITLE{} uses non-invasive, framework-level instrumentation and syscall-based causal analysis to cover all execution paths.

\item \textbf{Black-box Correlation.}
Systems like \textsc{Sanare} \cite{matos2021sanare} and \textsc{Rectify} \cite{matos2017rectify} treat the application as a black box, using machine learning to correlate high-level requests with low-level system events. Their accuracy degrades significantly under high concurrency due to the noisy and interleaved nature of system events. Furthermore, their file monitoring is often restricted to predefined directories. \TITLE{} achieves superior accuracy through deterministic partitioning and attribution, and its provenance analysis provides whole-system file tracking.

\item \textbf{System-wide Causal Tracking.} 
Approaches like \textsc{Retro} \cite{kim2010intrusion} build a global graph of system actions to support recovery. In a high-concurrency  web environment, this leads to dependency explosion\cite{lee2013high,ma2017mpi,alhanahnah2022autompi,yu2021alchemist,wang2023tesec,hassan2020omegalog}, creating a massive, tangled graph that is intractable to analyze. \TITLE{}'s core design overcomes this by first partitioning the global log into independent, request-specific causal chains, then using a targeted method to solve the most complex dependency (the database), thus avoiding the need to analyze a monolithic graph.

\item\textbf{Database-specific Recovery.} 
Work such as \textsc{Warp} \cite{chandra2011intrusion} focuses on fine-grained database recovery by rolling back and selectively replaying non-malicious transactions. However, these systems assume that the malicious transactions are already identified and do not address the challenge of attributing transactions to a specific web request in modern architectures with connection pooling. \TITLE's key contribution is solving this attribution problem with our spatiotemporal anchor method, providing the missing link that makes rollback-and-replay strategies like \textsc{Warp}'s practical in real-world settings.
\end{itemize}

\section{Conclusion}
\label{sec:conclusion}

This paper introduced \TITLE{}, a novel system for precise and comprehensive intrusion recovery in modern web applications. To overcome the dependency explosion problem in high-concurrency environments, \TITLE{} first employs deterministic request partitioning using framework-level delimiter logging to isolate the syscalls of individual requests. Following this partitioning, \TITLE{} addresses state modifications across both the file system and the database. For the file system, it constructs a syscall-based provenance graph to uncover all attack-induced modifications, including those from unexpected execution paths. For the database, it deconstructs the asynchronous interaction model using a spatiotemporal anchor, the combination of a network connection and its active time window, to deterministically attribute database operations to the originating request. Our evaluation on 10 web applications employing databases including \texttt{MySQL}, \texttt{PostgreSQL}, and \texttt{MongoDB} across 20 CVE-based attack scenarios demonstrates that \TITLE{} achieves 99.9\% accuracy under concurrency levels up to 150 connections. \TITLE{} successfully reverses attack-induced damage while preserving legitimate user data, with manageable performance overhead: 19.8\% average response latency increase, 17.8\% QPS reduction, and recovery throughput of 110.7 database operations per second and 27.2 affected files per second.

\nocite{du2017deeplog,liu2019log2vec,xie2019p-gaussian,milajerdi2019holmes,hassan2018towards,hossain2017sleuth}

\bibliographystyle{IEEEtran}

\bibliography{ancora}





\end{document}